\documentstyle[sprocl,epsfig]{article}

\def\Journal#1#2#3#4{{#1} {\bf #2}, #3 (#4)}

\def\NPB{{\em Nucl. Phys.} B}

\def\ZPC{{\em Z. Phys.} C}

\def\be{\begin{equation}}
\def\ee{\end{equation}}
\def\bea{\begin{eqnarray}}
\def\eea{\end{eqnarray}}

\newcommand{\xgo}{\mbox{$x_{\gamma}^{\rm obs}$}}

\begin{document}

\title{HIGH $E_T$ PHOTOPRODUCTION AND THE PHOTON STRUCTURE}

\author{M. WING \\(On behalf of the H1 and ZEUS collaborations)}

\address{McGill University, Physics Department, 3600 University Street, Montreal,\\ 
Canada, H3A 2T8 \\E-mail: wing@mail.desy.de} 


\maketitle\abstracts{Dijet photoproduction results are presented from the H1 and 
ZEUS collaborations and compared to next-to-leading order pQCD calculations.}


\section{Introduction}

The study of high transverse energy ($E_T$) dijet photoproduction provides both a 
test of pQCD and information on the structure of the photon. In events with jets of  
sufficiently high $E_T$, the contribution of non-perturbative effects, which are 
currently poorly understood, is suppressed. The effect of 
hadronisation is also reduced, allowing a more meaningful comparison of 
hadron-level data with parton-level next-to-leading-order (NLO) calculations. 
In the kinematic regime of the current measurements, the parton densities 
of the proton are well constrained, in contrast to the parton densities of the photon. 
The parameterisations 
currently fit $F_2^\gamma$ data from $e^+e^-$ experiments which do not well constrain 
the gluon density of the photon or the quark densities at high $x_\gamma$, 
the fraction of momentum of the photon carried by the struck parton. 

In this paper, events with an almost real photon (virtuality, $Q^2\approx0~{\rm GeV^2}$)
containing at least two jets reconstructed using the $k_T$ clustering 
algorithm~\cite{kt} are analysed. Measurements of events with a virtual 
photon and with jets of lower $E_T$ are discussed in detail 
elsewhere~\cite{steve,virtual}. The H1 and ZEUS collaborations have analysed similar 
amounts of luminosity; 36~${\rm pb^{-1}}$ and 38~${\rm pb^{-1}}$ respectively, 
allowing measurements up to $E_T^{\rm jet}~\sim~90$~GeV. Improvements in the 
understanding of pQCD for jet photoproduction has led to the agreement of many 
independent calculations to within $5-10\%$~\cite{zeus95,nlo}. 

\section{Results}

Cross sections as a function of the transverse momentum of the 
leading jet and the average transverse momentum of the two jets are shown in 
Fig.~\ref{fig:h12jets} compared to expectations from the {\sc Pythia} Monte 
Carlo~\cite{pythia} program and an NLO calculation. The kinematic regime is shown in  
Fig.~\ref{fig:h12jets} where the leading jet was required to have  
$P_T^{\rm jet1}~>~25$~GeV. The transverse momentum of the jet falls by three orders of 
magnitude up to values of $\sim$ 90 GeV. The cross section measurements are well 
described by the pQCD calculation in both shape and normalisation.

\centerline{\epsfig{figure=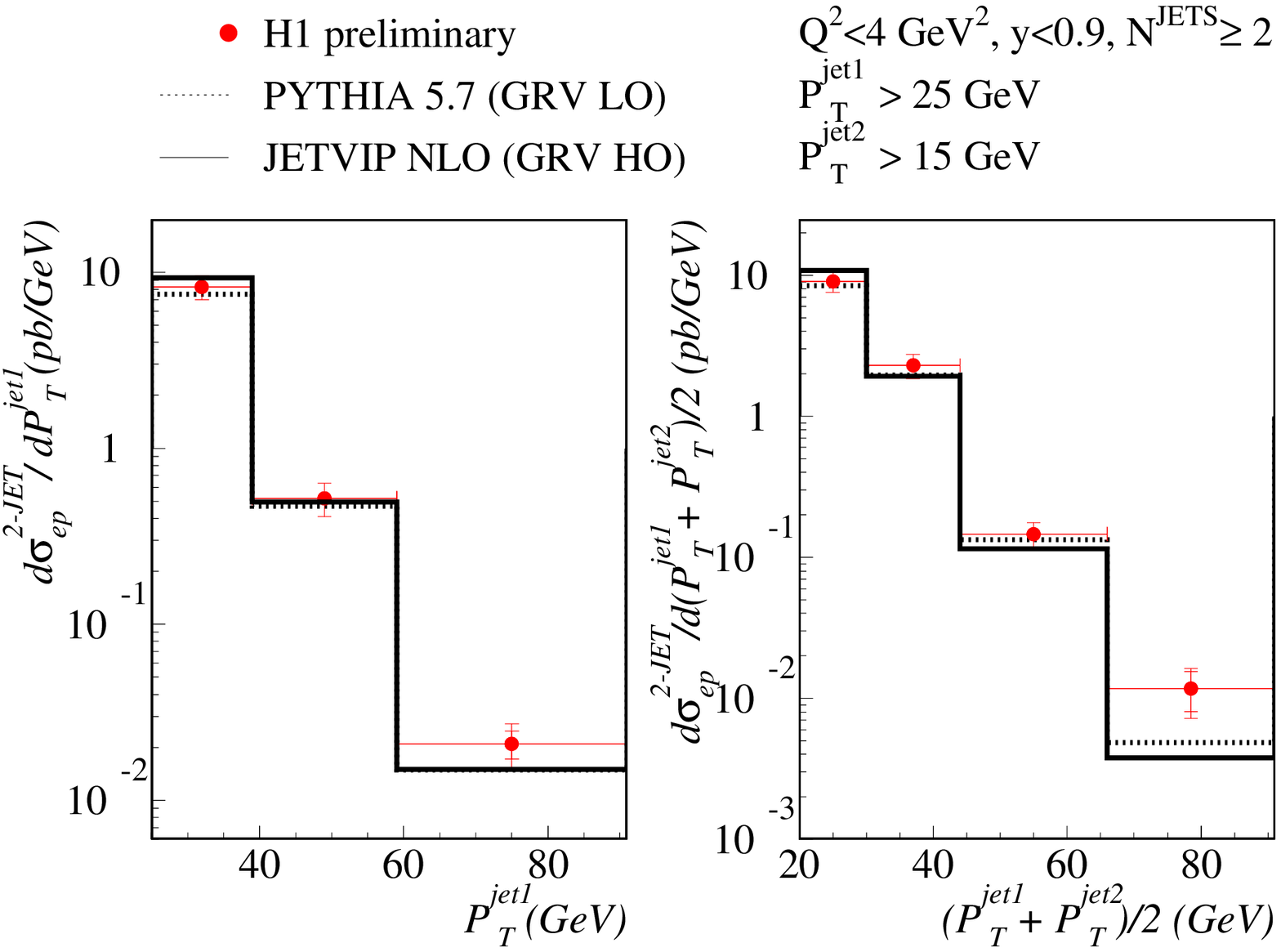,height=6.5cm}}
\begin{figure}[ht]
\vspace{-0.5cm}
\caption{Cross section measurements in transverse momentum of the leading jet, 
$P_T^{\rm jet1}$ and the average transverse momentum of the two leading jets, 
$(P_T^{\rm jet1}+P_T^{\rm jet1})/2$ compared to expectations.\label{fig:h12jets}}
\end{figure}

As the pQCD calculations describe the measurements at very high values of 
transverse energy, lowering the energy and considering other quantities provides 
a test of the current photon parton density functions (PDF's). It has 
already been demonstrated that at forward regions of pseudorapidity of the jet, 
$\eta^{\rm jet}$, the NLO calculations underestimate the data for all choices of 
photon PDF~\cite{zeus95,zeuseps}. Requiring that \xgo, the fraction of the 
photon's energy participating in the production of the two highest energy jets;
\begin{equation}
\xgo = \frac{\sum_{\rm jet1,2} E_T^{\rm jet} e^{-\eta^{\rm jet}}}{2yE_e},
\end{equation}
where $yE_e$ is the initial photon energy, satisfies $\xgo>0.75$, enhances the 
direct photon component of the cross section. 
The NLO calculations were shown to agree with the data in this region justifying 
the validity of the calculations and indicating that the discrepancy when no \xgo \ 
cut is applied arises due to the inadequacies of the photon PDF's. 

This was investigated further by considering the cross section in \xgo. The 
measured cross sections in \xgo \ 
for increasing slices in transverse energy of the leading jet are shown in 
Fig.~\ref{fig:xgamma} compared to an NLO calculation using the AFG-HO~\cite{afg} 
photon PDF. The measured data clearly lie up to $50-60\%$ above the pQCD calculation 
for low values of \xgo \ and the trend is maintained up to high values of 
$E_T^{\rm jet1}$.

\centerline{\epsfig{figure=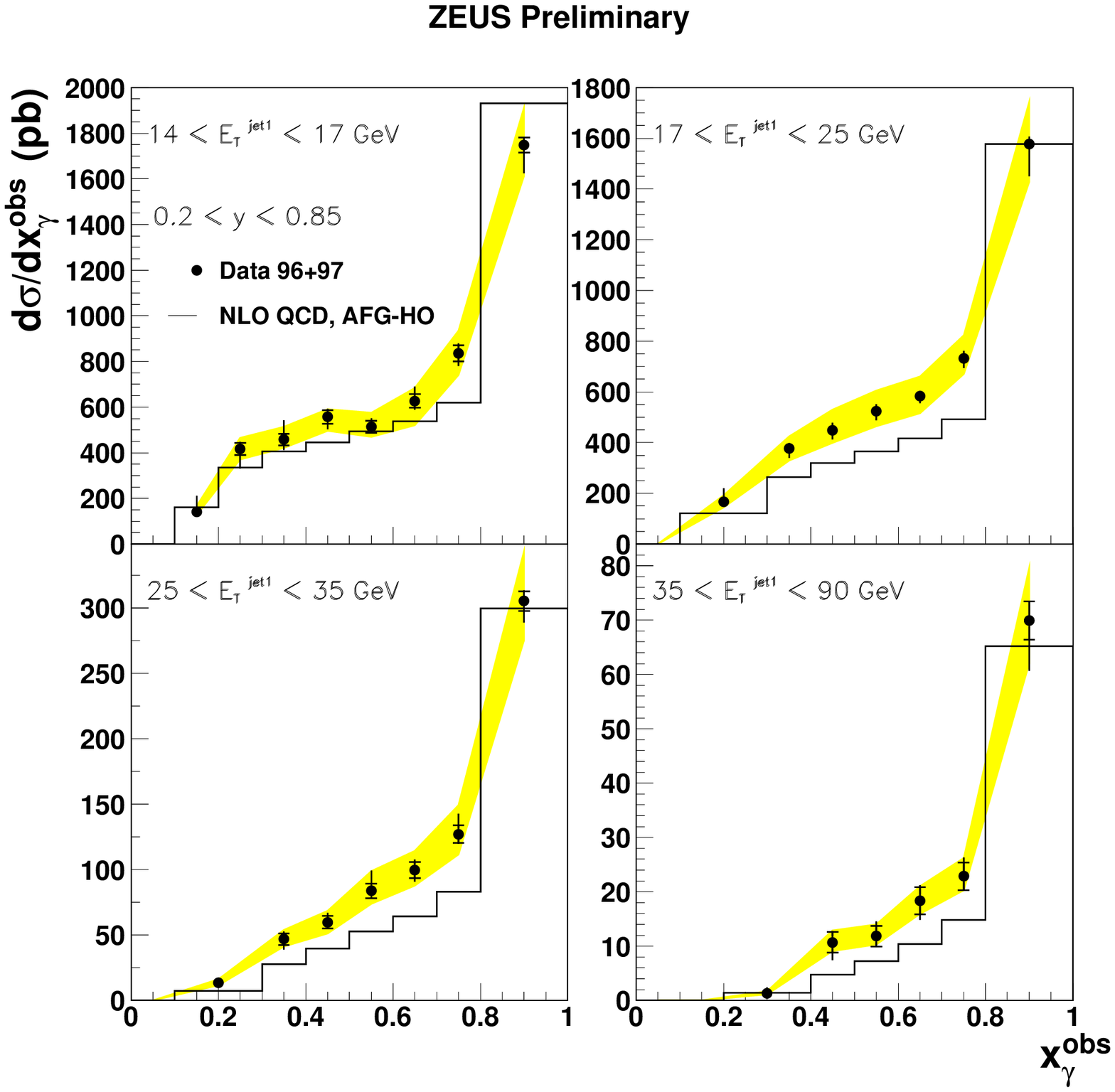,height=6.5cm,width=8cm}}
\begin{figure}[htp]
\vspace{-0.5cm}
\caption{Cross section measurements in \xgo \ in slices of $E_T^{\rm jet1}$.
The jets were restricted to $-1<\eta^{\rm jet}<2$ and the events to the region  
$Q^2<1~{\rm GeV^2}$ and $0.2<y<0.8$. The band around the points displays the 
error due to the uncertainty associated with the calorimeter energy scale. 
\label{fig:xgamma}}
\end{figure}

Having established a region of phase space in which the data is not described by the 
pQCD calculations, further studies were made to ascertain 
if the problem lies with the calculation or the photon PDF. This was done by considering 
the cross section in $|{\rm cos}\theta^*|$, the angle between the dijet and beam 
axes in the dijet centre-of-mass system. This measurement directly 
tests the parton-parton dynamics of the hard sub-process and was performed  
in two regions of \xgo \ ($\xgo<0.75$ and $\xgo>0.75$). Additional requirements were 
made to remove the biases in the distribution due to the $E_T^{\rm jet}$ and 
$\eta^{\rm jet}$ cuts. Cuts on the dijet invariant mass, $M_{jj}~>~39$~GeV and the 
average pseudorapidity of the jets, $0~<~\bar{\eta}~<~1$, were made.

\begin{figure}[htp]
\centerline{\epsfig{figure=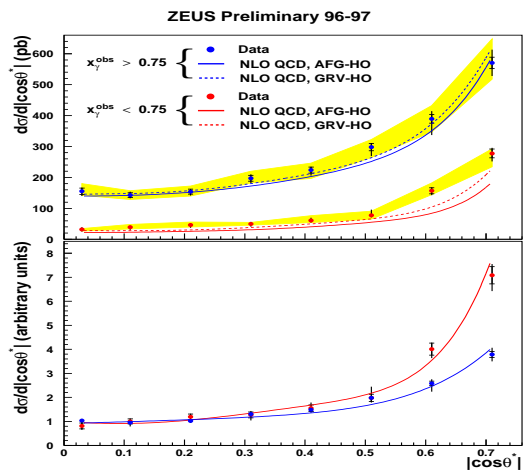,height=6.5cm,width=7cm}}
\vspace{-0.5cm}
\caption{Cross section in $|{\rm cos}\theta^*|$ compared to NLO cross section 
predictions and shape where the distributions are normalised 
such that the integral of the first three bins is equal. \label{fig:cos}}
\end{figure}

The cross section is shown in Fig.~\ref{fig:cos} compared to the absolute prediction 
and the shape of the NLO calculation. As has already been observed, the higher \xgo \ 
region is well described and the lower \xgo \ region poorly described in normalisation. 
However, the shape of the cross section for $\xgo<0.75$ is well described by 
the NLO calculation indicating that the calculation describes 
the parton-parton dynamics of the hard sub-process.

\section{Conclusions}

Dijet photoproduction has been measured up to $E_T^{\rm jet} \sim$ 90 GeV 
and compared to pQCD calculations. Clear evidence is seen that the calculations 
give a good description of the scattering process. However, inadequacies in the 
current parameterisations of the photon PDFs are evident. The use of the data 
presented here in future fits would greatly improve our understanding of the 
photon PDF.

\end{document}